\documentclass[fleqn,reqno]{amsart}

\usepackage{graphicx,amscd,amsmath,amssymb,verbatim}
\usepackage[dvips]{hyperref}
\usepackage[TS1,OT1,T1]{fontenc}

\begin{document}

\title[Noninertial relativity consistent with Heisenberg algebra]{Noninertial Relativity group with invariant Minkowski metric consistent with Heisenberg quantum commutation relations}
\author{Stephen G. Low}
\address{Austin, Texas }
\email{Stephen.Low@alumni.utexas.net}
\date{\today}
\maketitle
\begin{abstract}

The inhomogeneous Lorentz group defines the transformations between
inertial states and special relativistic quantum mechanics is defined
in terms of its projective representations. Special relativity does
not address how noninertial states are related. If the noninertial
system is due to gravity, general relativity resolves this through
a curved manifold where particles under the action of gravity follow
geodesics that are locally inertial trajectories. However, general
relativity also does not address the issue of how the states of
noninertial particles on a flat space due to a force other than
gravity are related.\ \ We study this by starting with a quantum
system with physical observables of position, time, energy and momentum
that are the Hermitian representation of the generators of the algebra
of the Weyl-Heisenberg group.\ \ We require that this is true for
any states related by the projective representation of the relativity
group. We show that this results in a consistency condition that
requires the relativity group to be a subgroup of the group of automorphisms
of the Weyl-Heisenberg algebra and consider the relativity groups
that also leaves invariant a Minkowski line element. This defines
the expected noninertial relativistic transformations and that have
the expected classical limit as $c\rightarrow \infty $ .\ \ In a
companion paper, a quantum mechanics for this noninertial relativity
group is formulated in terms of the projective representations of
the inhomogeneous group using the same approach as for special relativistic
quantum mechanics.
\end{abstract}

\section{Introduction}

The inhomogeneous Lorentz group defines the relation between inertial
states.\ \ Clocks locally at rest to a state are related to the
clocks of other inertial observers through the Minkowski proper
time line element.\ \ Quantum states are rays in a Hilbert space
and therefore inertial states are related through the projective
representation of the inhomogeneous Lorentz group. As projective
representations are equivalent to the unitary representations of
the central extension,\ \ these are the unitary representations
of the Poincar\'e group that is the cover of the inhomogeneous Lorentz
group \cite{wigner},\cite{bargmann},\cite{mackey2},\cite{Weinberg1}.

The equivalence principle of general relativity enables the noninertial
frames of a particle accelerating under gravity to be understood
as locally inertial frames on a curved manifold. Particles under
gravity follow geodesics and neighboring locally inertial frames
are related by the connection. The clock locally at inertial rest
is related to the local clocks of other neighboring observers in
the gravitating system through the Riemannian proper time\ \ line
element. 

Neither general relativity nor special relativity addresses the
issue of noninertial states that are not due to gravity, but rather
one of the other forces, and therefore the underlying manifold is
flat. Consider for example an electron in a region that gravity
is negligible that encounters an electromagnetic field and therefore
perturbs to a noninertial trajectory and is observed also by an
observer in an apparent inertial frame.\ \ How is the clock of this
noninertial state related to the clocks of other observers?

We hypothesize that the noninertial relativity group relating these
states is the most general group consistent with the requirements
that 

1) the Heisenberg uncertainty principle holds in the noninertial
as well as inertial states

2) the proper time given by the Minkowski line element that is invariant
in noninertial states

To make this more precise, we first consider a quantum system in
which the position, momentum, energy and time degrees of freedom
are represented by the Hermitian representation of the algebra of
the Weyl-Heisenberg group $\mathcal{H}( n+1) $ where the number
of spacial dimensions is $n=3$. The requirement that the algebra
transforms into itself under the action of the relativity group
means that the relativity group a subgroup of the automorphism group
of the Weyl-Heisenberg algebra.\ \ This automorphism group is \cite{folland,Low7}
\begin{equation}
{\mathcal{A}ut}_{\mathcal{H}}\simeq \mathbb{Z}_{2}\otimes _{s}\mathcal{D}\otimes
_{s}\overline{\mathcal{H}\mathcal{S}p}( 2n+2) ,%
\label{nirg: autH definition}
\end{equation}

\noindent where\ \ $\mathcal{H}\mathcal{S}p( 2n+2) \simeq \mathcal{S}p(
2n+2) \otimes _{s}\mathcal{H}( n+1) $. $\mathbb{Z}_{2}$ is the 2
element discrete group, $\mathcal{D}$ is the abelian group isomorphic
to the reals under multiplication, $\mathcal{S}p( 2n+2) $ is the
symplectic group and $\mathcal{H}( n+1) $ is the Weyl-Heisenberg
group. 

The Minkowski line element is\ \ $d \tau ^{2 }=d t^{2}-\frac{1}{c^{2}}d
q^{2}$. This is an invariant for states that are inertially related
and the second assertion is that this continues to be true for general
noninertial states.\ \ \ \ 

We will show that the homogeneous relativity group that is a subgroup
of the automorphism group of the Weyl-Heisenberg group\ \ that leaves
the Minkowski line element invariant is
\begin{equation}
\mathcal{U}b( 1,n) \simeq \mathcal{O}( 1,n) \otimes _{s}\mathcal{A}(
m) ,
\end{equation}

\noindent where $m=(n+1)(n+2)/2$ and $\mathcal{A}( m) $ is the abelian
group isomorphic to $\mathbb{R}^{m}$ under addition.\ \ The additional
generators of the abelian group behave as a power-force stress tensor
that is the proper time derivative of the energy-momentum stress
tensor. We show that this leads to expected relativistic results
in transforming to noninertial states \cite{Low5}.\ \ 

This relativistic theory must lead to expected classical results
in the limit $c\rightarrow \infty $ where the Minkowski line element
reduces to the invariant Newtonian time line element $d t^{2}$.
We have previously studied the most general group that leaves invariant
the Newtonian time line element $d t^{2}$ that is a subgroup of
the automorphisms of the Weyl-Heisenberg group. This results in
a group that leads directly to Hamilton's equations and, with the\ \ additional
requirement of orthonormal position frames,\ \ \ describes the Hamilton
relativity group for noninertial transformations in a classical
context \cite{Low7}, \cite{Low8}. 

\section{Consistency between a relativity group and quantum mechanics}

States in quantum mechanics are represented by rays $\Psi $ in a
Hilbert space $\text{\boldmath $\mathrm{H}$}$ that are the equivalence
class of states in the Hilbert space related by a phase
\begin{equation}
\Psi \simeq \left\{  e ^{i \omega }\left| \psi \right\rangle  |\omega
\in \mathbb{R}\right\} ,
\end{equation}

\noindent where $|\psi \rangle \in \text{\boldmath $\mathrm{H}$}$.\ \ A
relativity group $g\in \mathcal{G}$ acts on the states through a
projective representation $\pi $, $\tilde{\Psi }=\pi ( g) \Psi $,\ \ with
the property that
\begin{equation}
\pi ( \tilde{g}\cdot g) = e ^{i \omega ( \tilde{g},g) }\pi ( \tilde{g})
\pi ( g) , \omega ( \tilde{g},g) \in \mathbb{R}\text{}.
\end{equation}

\noindent Projective representations are equivalent to the unitary
representations $\varrho $ of the central extension $\check{\mathcal{G}}$
of the group $\mathcal{G}$ \cite{bargmann}, \cite{mackey2} that
act on the states as
\begin{equation}
\left. \tilde{\left. \left| \psi \right. \right\rangle  }=\varrho
( g) \left| \psi \right. \right\rangle  ,\ \ g\in \check{\mathcal{G}},\ \ \ \ \ \ \left.
\left| \psi \right. \right\rangle  \in {\text{\boldmath $\mathrm{H}$}}^{\varrho
}.
\end{equation}

\noindent The Hilbert space is determined by the unitary representation
$\varrho $ and so we label it as ${\text{\boldmath $\mathrm{H}$}}^{\varrho
}$.\ \ Observables corresponding to a the relativity group $\mathcal{G}$
are represented by the Hermitian representations $\varrho ^{\prime
}$ of the algebra of a group $ \check{\mathcal{G}}$ , $\hat{Z}=\varrho
^{\prime }( Z) $. The action of the group element $g\in \check{\mathcal{G}}$
on these observables is 
\begin{equation}
\hat{\tilde{Z}}\tilde{\left| \psi \right\rangle  }=\varrho ( g )
\hat{Z}\left| \psi \right\rangle  =\varrho ( g ) \hat{Z}{\varrho
( g ) }^{-1}\varrho ( g ) \left| \psi \right\rangle  =\varrho (
g ) \hat{Z}{\varrho ( g ) }^{-1}\tilde{\left| \psi \right\rangle
}%
\label{nirg: projective tx of algebra}
\end{equation}

\noindent and so 
\begin{equation}
\hat{\tilde{Z}}=\varrho ^{\prime }( \tilde{Z}) =\varrho ( g ) \hat{Z}{\varrho
( g ) }^{-1}=\varrho ( g ) \varrho ^{\prime }( Z) {\varrho ( g )
}^{-1}=\varrho ^{\prime }( g Z g^{-1}) .
\end{equation}

\noindent Therefore, if the representation $\varrho $ is faithful,
we have that
\begin{equation}
\tilde{Z}=g Z g^{-1}%
\label{nirg: transformation of an observable}
\end{equation}

\noindent and otherwise this is an equivalence up to the kernel
of the homomorphism. 

 Position, momentum, energy and time observables are the Hermitian
representation of the algebra of the Weyl-Heisenberg group $\mathcal{H}(
n+1) $\ \ with a general element given by $\begin{array}{rl}
 Z & =z^{\alpha }Z_{\alpha }
\end{array}$, $\alpha =1,.. 2n+2$ where $\{z^{\alpha }\}\in \mathbb{P}\simeq
\mathbb{R}^{2n+2}$ and $Z_{\alpha }$ are a dimensionless basis for
the Weyl-Heisenberg algebra that satisfy the commutation relations\ \ 
\begin{equation}
\left[ Z_{\alpha },Z_{\beta }\right] = \zeta _{\alpha ,\beta }I,
\end{equation}

\noindent and $\zeta _{\alpha ,\beta }$ are the components of a
symplectic metric.\ \ The Hermitian representation of the algebra
satisfies 
\begin{equation}
\left[ {\hat{Z}}_{\alpha },{\hat{Z}}_{\beta }\right] =i \zeta _{\alpha
,\beta }\hat{I},
\end{equation}

\noindent where $\hat{Z_{\alpha }}=\varrho ^{\prime }( Z_{\alpha
}) $ and $\hat{I}=\varrho ^{\prime }( I) $ is the unit operator
on the Hilbert space. Set $\{\hat{Z_{\alpha }}\}=\{{\hat{P}}_{i},{\hat{Q}}_{i},\hat{E},\hat{T}\}$
with $i=1,..n$.\ \ \ These are the familiar Hermitian representations
that in a basis with position and time diagonal, are 
\begin{equation}
\begin{array}{ll}
 \left. \left. \left\langle  q,t\right. \right| {\hat{Q}}_{i}\left|
\psi \right. \right\rangle  =q^{i}\psi ( q,t) , & \left. \left.
\left\langle  q,t\right. \right| \hat{T}\left| \psi \right. \right\rangle
=t \psi ( q,t) , \\
 \left. \left. \left\langle  q,t\right. \right| {\hat{P}}_{i}\left|
\psi \right. \right\rangle  =i \hbar \partial \psi ( q,t) /\partial
q^{i}, & \left. \left. \left\langle  q,t\right. \right| \hat{E}\left|
\psi \right. \right\rangle  =-i \hbar \partial \psi ( q,t) /\partial
t. 
\end{array}
\end{equation}

Bases that diagonalize other commuting sets $\psi ( p,t) =\langle
q,t|\psi \rangle $, $\psi ( p,e) =\langle q,e|\psi \rangle $, $\psi
( q,e) =\langle q,e|\psi \rangle $ and the corresponding representations
of the operators\ \ $\{{\hat{P}}_{i},{\hat{Q}}_{i},\hat{E},\hat{T}\}$
in these bases are equally valid \cite{dirac}.\ \ Generally, our
bias is to diagonalize the position time basis $\psi ( q,t) =\langle
q,t|\psi \rangle $.

The basic physical assumption is that the Heisenberg commutation
relations are satisfied by any basis related by a relativity group
$ \mathcal{G}$.\ \ That is,\ \ position, momentum, energy and time
observables satisfying the Heisenberg quantum commutation relations
will also satisfy the Heisenberg quantum commutation relations for
any states related by the projective representations of the relativity
group (6). This implies using (8) that if $\{Z_{\alpha },I\}$ are
a basis of the Weyl-Heisenberg algebra, then $\{{\tilde{Z}}_{\alpha
},\tilde{I}\}$ are also a basis of the Weyl-Heisenberg algebra where
\begin{equation}
\tilde{Z_{\alpha }}=g Z_{\alpha } g^{-1}, \tilde{I}=g I g^{-1}=I.
\end{equation}

\noindent and $g\in \check{\mathcal{G}}$ and $\varrho$ is a faithful
representation.\ \ The maximal group for which this property is
true is the automorphism group of the Weyl-Heisenberg group.\ \ This
results in basic consistency condition that the central extension
$\check{\mathcal{G}}$ of the relativity group $\mathcal{G}$ must
be a subgroup of the automorphism group of the Weyl-Heisenberg algebra
${\mathcal{A}ut}_{\mathcal{H}( n+1) }$,
\begin{equation}
\check{\mathcal{G}}\subseteq {\mathcal{A}ut}_{\mathcal{H}( n+1)
}.
\end{equation}

 The automorphism group of the Weyl-Heisenberg group is \cite{folland}
\begin{equation}
{\mathcal{A}ut}_{\mathcal{H}( n+1) }={\overline{\mathcal{O}\mathcal{A}ut}}_{\mathcal{H}(
n+1) }\otimes _{s}\mathcal{H}( n+1) 
\end{equation}

\noindent where the Heisenberg group itself are the inner automorphisms.\ \ The
outer automorphisms are 
\begin{equation}
{\mathcal{O}\mathcal{A}ut}_{\mathcal{H}( n+1) }\simeq \mathbb{Z}_{2}\otimes
\mathcal{D}\otimes \mathcal{S}p( 2n+2) .
\end{equation}

\noindent The matrix realization of this group and the group properties
are given in Appendix A.\ \ The central extension is
\begin{equation}
{\check{\mathcal{O}\mathcal{A}ut}}_{\mathcal{H}( n+1) }\simeq {\overline{\mathcal{O}\mathcal{A}ut}}_{\mathcal{H}(
n+1) }\simeq \mathbb{Z}_{2}\otimes \mathcal{D}\otimes \overline{\mathcal{S}p}(
2n+2) .
\end{equation}

\noindent Therefore, the relativity group may always be written
as 
\begin{equation}
\check{\mathcal{G}}\subseteq \check{\mathcal{K}}\otimes _{s}\mathcal{N}
\end{equation}

\noindent where $\mathcal{K}$ is the homogeneous relativity group
that is a subgroup of the outer automorphisms, $\mathcal{K}\subseteq
{\mathcal{O}\mathcal{A}ut}_{\mathcal{H}( n+1) }$ and $\mathcal{N}\subseteq
\mathcal{H}( n+1) $ and $\check{\mathcal{G}}\subseteq {\mathcal{O}\mathcal{A}ut}_{\mathcal{H}(
n+1) }$.

\section{Homogeneous relativity group }

We determine in this section the homogeneous relativity group for
noninertial frames that satisfies two conditions. 

1) It leaves invariant the Minkowski proper time line element. The
line element that is the invariant also of the inertial frames of
special relativity is valid also for the noninertial case. 

2) It is a subgroup of the automorphism group of the Weyl-Heisenberg
group. Therefore, the Heisenberg commutation relations hold in all
states related by this relativity group and therefore from the previous
section must be a subgroup of ${\mathcal{O}\mathcal{A}ut}_{\mathcal{H}(
n+1) }$.

We name the relativity group that satisfies these two conditions
$\mathcal{U}b( 1,n) $ and use it to study relativistic noninertial
transformations. 

The group $\mathcal{U}b( 1,n) $ is dependent on the scale $c$.\ \ We
show a homomorphism parameterized by $c$ satisfies the conditions
to define a In\"on\"u-Wigner contraction \cite{Glimore2}.\ \ This
contraction results in the\ \ Hamilton group that we have previously
shown is the relativity group for noninertial frames in the classical
($c\rightarrow \infty )$ context.\ \ 

\subsection{The group $\mathcal{U}b( 1,n) $ and its algebra}

The postulates of special relativity requires the invariance of
the Minkowski proper time line element
\begin{equation}
d \tau ^{2} = \eta _{a,b}d x^{a}d x^{b}
\end{equation}

\noindent with $a,b..=0,..n$ and $\eta $ is the diagonal matrix\ \ $\eta
=[\eta _{a,b}]=\mathrm{diag}\{-1,1,...1\}$ and units where $c=1$.

Consider the $2n+2$ dimensional time, position, energy, momentum
space\ \ $\mathbb{P}\simeq \mathbb{R}^{2n+2}$ with coordinates $\{z^{\alpha
}\}=\{x^{a},p^{a}\}$ where $\alpha ,\beta =1,...2n+2$, $a,b=0,1..n$.\ \ The
Minkowski metric may be considered to be a degenerate line element
on the cotangent space $T_{z}^{*}\mathbb{P}$ 
\begin{equation}
d \tau ^{2} = {\tilde{\eta }}_{\alpha ,\beta }d z^{\alpha }d z^{\beta
}%
\label{nirg: Minkowski line element}
\end{equation}

\noindent where ${\tilde{\eta }}_{\alpha ,\beta }$ are the components
of the $(2n+2)\times (2n+2)$ dimensional matrix $\tilde{\eta }$
\[
\tilde{\eta }= \left[ {\tilde{\eta }}_{\alpha ,\beta }\right] =\left(
\begin{array}{ll}
 \left[ \eta _{a,b}\right]  & 0 \\
 0 & 0
\end{array}\right) .
\]

The group $\mathcal{G}\mathcal{L}( 2n+2,\mathbb{R}) $ of nonsingular
$(2n+2)\times (2n+2)$ matrices acts naturally on the cotangent space
$T_{z}^{*}\mathbb{P}$ with basis $\{d z^{\alpha }|_{z}\}$.\ \ Elements
$\Gamma $ of the subgroup $\mathcal{O}b( 1,n)  \subset \mathcal{G}\mathcal{L}(
2n+2,\mathbb{R}) $ that leave invariant the degenerate line element\ \ (19)
satisfy\ \ \cite{Glimore2}
\begin{equation}
{}^{t}\Gamma  \tilde{\eta } \Gamma =\tilde{\eta }.%
\label{nirg: orthogonal line element condition}
\end{equation}

\noindent $\Gamma $ may be written in terms of\ \ $(n+1)\times (n+1)$
submatrices as $\Gamma =(\begin{array}{ll}
 \Lambda  & B \\
 \Xi  & A
\end{array})$ and therefore using (20)
\begin{equation}
\begin{array}{rl}
 \left( \begin{array}{ll}
 \eta  & 0 \\
 0 & 0
\end{array}\right)  & =\left( \begin{array}{ll}
 {}^{t}\Lambda  & {}^{t}\Xi  \\
 {}^{t}B & {}^{t}A
\end{array}\right) \left( \begin{array}{ll}
 \eta  & 0 \\
 0 & 0
\end{array}\right) \left( \begin{array}{ll}
 \Lambda  & B \\
 \Xi  & A
\end{array}\right)  \\
  & =\left( \begin{array}{ll}
 {}^{t}\Lambda  \eta  \Lambda  & {}^{t}\Lambda  \eta  B \\
 {}^{t}B \eta  \Lambda  & {}^{t}B \eta  B
\end{array}\right) .
\end{array}%
\label{nirg: group leaving minkowski invariant}
\end{equation}

\noindent It follows immediately that $B=0$ and $\Lambda \in \mathcal{O}(
1,n) $ and as the $\det  \Gamma =\mathrm{det\Lambda } \det  A $,
$\det  A\neq 0$.\ \ \ Therefore, 
\begin{equation}
\mathcal{O}b( 1,n)  \simeq \left( \mathcal{O}( 1,n) \otimes \mathcal{G}\mathcal{L}(
n+1,\mathbb{R}) \right) \otimes _{s}\mathcal{A}( {\left( n+1\right)
}^{2}) 
\end{equation}

\noindent with elements $\Gamma $\ \ and $\Gamma ^{-1}$ of the form
\begin{equation}
\Gamma =\left( \begin{array}{ll}
 \Lambda  & 0 \\
 \Xi  & A
\end{array}\right) ,\ \ \ \ \Gamma ^{-1}=\left( \begin{array}{ll}
 \Lambda ^{-1} & 0 \\
 -A^{-1} \Xi  \Lambda ^{-1} & A^{-1}
\end{array}\right) .%
\label{nirg: Ob element and inverse}
\end{equation}

The homogeneous relativity group $\mathcal{U}b( 1,n) $ must be a
subgroup of the group of outer automorphisms ${\mathcal{O}\mathcal{A}ut}_{\mathcal{H}(
n+1) }$ and also the group $\mathcal{O}b( 1,n) $ that leaves the
degenerate line element invariant,\ \ 
\begin{equation}
\mathcal{U}b( 1,n) ={\mathcal{O}\mathcal{A}ut}_{\mathcal{H}( n+1)
}\cap \mathcal{O}b( 1,n) .
\end{equation}

The elements of the outer automorphism group are of the form $\Delta
\Sigma $\ \ where $\Delta \in \mathbb{Z}_{2}\otimes \mathcal{D}$
and $\Sigma \in \mathcal{S}p( 2n+2) $ as given in Appendix A. The
symplectic matrices satisfy the condition ${}^{t}\Sigma  \zeta 
\Sigma  =\zeta $ and so $\Sigma ^{-1}=-\zeta {}^{t}\Sigma  \zeta
$ with $\zeta =(\begin{array}{ll}
 0 & \eta  \\
 -\eta  & 0
\end{array})$. This may also be written in terms of $(n+1)\times
(n+1)$ submatrices $\Sigma _{\mu ,\nu }$\ \ $\mu ,\nu =1,2$\ \ with
the matrix and inverse having the form
\begin{equation}
\Sigma =\left( \begin{array}{ll}
  \Sigma _{1,1} & \Sigma _{1,2} \\
 \Sigma _{2,1} & \Sigma _{2,2}
\end{array}\right) ,\ \ \ \Sigma ^{-1}=\left( \begin{array}{ll}
 \eta {}^{t}\Sigma _{2,2}\eta  & -\eta  {}^{t}\Sigma _{1,2}\eta
\\
 {}-{\eta  }^{t}\Sigma _{2,1}\eta  & \eta  {}^{t}\Sigma _{1,1}\eta
\end{array}\right) .
\end{equation}

Therefore, if $\Gamma $ in (23) is a subgroup of the outer automorphism
group, we have $\Sigma =\Delta ^{-1}\Gamma $ and so 
\begin{equation}
\left( \begin{array}{ll}
  \Sigma _{1,1} & \Sigma _{1,2} \\
 \Sigma _{2,1} & \Sigma _{2,2}
\end{array}\right) =\Delta ^{-1}\left( \begin{array}{ll}
 \eta  \Lambda  \eta  & 0 \\
 -\eta  \Xi  \eta  & \eta A\eta 
\end{array}\right) 
\end{equation}

\noindent and\ \ as $\Gamma ^{-1 }=\Sigma ^{-1}\Delta ^{-1}$, we
also have 
\begin{equation}
\left( \begin{array}{ll}
 {}^{t}\Sigma _{2,2} & {}^{t}\Sigma _{1,2} \\
 {}^{t}\Sigma _{2,1} &  {}^{t}\Sigma _{1,1}
\end{array}\right) =\Delta ^{-1}\left( \begin{array}{ll}
 {}{\eta  }^{t}A \eta  & 0 \\
 {}{\eta  }^{t}\Xi  \eta  & \eta  {}^{t}\Lambda  \eta 
\end{array}\right)  \Delta ^{-1}=\Delta ^{-2}\left( \begin{array}{ll}
 {}{\eta  }^{t}A \eta  & 0 \\
 {}{\eta  }^{t}\Xi  \eta  & \eta  {}^{t}\Lambda  \eta 
\end{array}\right) .
\end{equation}

\noindent Finally, equating to the inverse $\Gamma ^{-1}$ previously
calculated in (23) 
\begin{equation}
\ \ \ \Gamma ^{-1}=\left( \begin{array}{ll}
 \Lambda ^{-1} & 0 \\
 -\Lambda ^{-1} \Xi  A^{-1} & A^{-1}
\end{array}\right) =\Delta ^{-2}\left( \begin{array}{ll}
 {}{\eta  }^{t}A \eta  & 0 \\
 {}-{\eta  }^{t}\Xi  \eta  & \eta  {}^{t}\Lambda  \eta 
\end{array}\right) %
\label{nirg: Ob element and inverse}
\end{equation}

\noindent from which it follows that $\Lambda ^{-1}=\Delta ^{-2}{}\eta
\ \ {}^{t}A \eta $ and $A^{-1}=\Delta ^{-2}\eta  {}^{t}\Lambda 
\eta $.\ \ This has a solution if and only if $\Delta =\pm I_{n}\in
\mathbb{Z}_{2}\subset \mathcal{D}$ and ${}^{t}A=\eta  \Lambda ^{-1}\eta
$.\ \ Noting that $\Lambda ^{-1}=\eta {}^{t}\Lambda  \eta $ this
gives ${}^{t}A={}^{t}\Lambda $ and therefore $A=\Lambda $. Finally,
\begin{equation}
{}^{t}\Xi =\eta  \Lambda ^{-1} \Xi  \Lambda ^{-1}\eta = {}^{t}\Lambda
\eta  \Xi  \eta \ \ {}^{t}\Lambda .
\end{equation}

 Thus elements of\ \ $\mathcal{U}b( 1,n) $ have the form 
\begin{equation}
\Gamma ( \Lambda ,\Xi ) =\left( \begin{array}{ll}
 \Lambda  & 0 \\
 \Xi  & \Lambda 
\end{array}\right) .
\end{equation}

\noindent In this expression, $\Lambda \in \mathcal{O}( 1,n) $.
The group multiplication and inverse of $\mathcal{U}b( 1,n) $ are
\begin{gather}
\begin{array}{rl}
 \Gamma ( \Lambda ,\Xi )  & =\Gamma ( \Lambda ^{\prime },\Xi ^{\prime
}) \Gamma ( \Lambda ^{{\prime\prime}},\Xi ^{{\prime\prime}})  \\
  & =\Gamma ( \Lambda ^{\prime } \Lambda ^{{\prime\prime}},\Xi ^{\prime
}\Lambda ^{{\prime\prime}}+{\Lambda  }^{\prime }\Xi ^{{\prime\prime}})
\end{array}, %
\label{nirg: Ub group transformation}
\end{gather}
\begin{equation}
{\Gamma ( \Lambda ,\Xi ) }^{-1}= \Gamma ( \Lambda ^{-1},-{\Lambda
}^{-1}\Xi  \Lambda ^{-1}) .
\end{equation}

The Lorentz group is the subgroup $\Gamma ( \Lambda ,0) $.\ \ The
matrix components of the Lorentz matrices may be given as the usual
expressions in regular and hyperbolic trigonometry terms of the
rotation angles and boost angles.

The elements $\Gamma ( I_{n}, \Xi  ) $ define an abelian normal
subgroup with group multiplication,\ \ inverse and automorphisms
given by 
\begin{gather}
\begin{array}{l}
 \Gamma ( I_{n},\Xi ^{\prime }) \Gamma ( I_{n},\Xi ^{{\prime\prime}})
=\Gamma ( I_{n},\Xi ^{\prime }+\Xi ^{{\prime\prime}}) , \\
 {\Gamma ( I_{n},\Xi ) }^{-1}= \Gamma ( I_{n},- \Xi  ) ,
\end{array}
\end{gather}
\begin{equation}
\begin{array}{rl}
 \Gamma ( \Lambda ^{\prime },\Xi ^{\prime }) \Gamma ( I_{n},\Xi
) {\Gamma ( \Lambda ^{\prime },\Xi ^{\prime }) }^{-1} & =\Gamma
( I_{n},\Lambda ^{\prime } \Xi  {\Lambda ^{\prime }}^{-1}) 
\end{array}.%
\label{nirg: inner automoprhisms of Ub}
\end{equation}

Also, for this subgroup ${}^{t}\Xi =\eta  \Xi  \eta $ and the matrix
components of $\Xi $ are the $(n+1)(n+2)/2$ real parameters $\xi
_{b}^{a}=\eta ^{a,d}\eta _{b,c}\xi _{d}^{c}$.\ \ Therefore $\Gamma
( I_{n},M) \in \mathcal{A}( (n+1)(n+2)/2) $. Consequently the full
group is 
\begin{equation}
\mathcal{U}b( 1,n)  \simeq \mathcal{O}( 1,n) \otimes _{s}\mathcal{A}(
\left( n+1\right) \left( n+2\right) /2) .%
\label{nirg: Ub definition}
\end{equation}

 It can be shown that it does not admit an algebraic central extension
and therefore the central extension of this group is simply its
cover
\begin{equation}
\overline{\mathcal{U}b}( 1,n) \simeq \overline{\mathcal{O}}( 1,n)
\otimes _{s}\mathcal{A}( \left( n+1\right) \left( n+2\right) /2)
.%
\label{nirg: definition of Ub}
\end{equation}

A general element of the algebra of $\mathcal{U}b( 1,n) $ is $Z=\lambda
^{a,b}L_{a,b}+\xi ^{a,b}M_{a,b}$ . Note that as $\xi ^{a,b}=\xi
^{b,a}$, that $M_{a,b}=M_{b,a}$.\ \ The\ \ Lie algebra relations
may be directly computed to be
\begin{equation}
\begin{array}{l}
 \left[ L_{a,b},L_{c,d}\right] =-L_{b,d} \eta _{a,c}+L_{b,c} \eta
_{a,d}+L_{a,d} \eta _{b,c}-L_{a,c} \eta _{b,d}, \\
 \left[ L_{a,b},M_{c,d}\right] =-M_{b,d} \eta _{a,c}-M_{b,c} \eta
_{a,d}+M_{a,d} \eta _{b,c}+M_{a,c} \eta _{b,d}, \\
 \left[ M_{a,b},M_{c,d}\right] =0.
\end{array}%
\label{nirg: Ub algebra}
\end{equation}

 The $M_{a,b}$ abelian generators transform as a symmetric $(0,2)$
tensor under the Lorentz generators $L_{a,b}$ .

Returning to the group, the transformation equations are\ \ $d \tilde{z}
=\Gamma  d z$.\ \ Using the definition of\ \ $\Gamma $ in (35)\ \ results
in
\begin{equation}
\begin{array}{l}
 d \tilde{x} = \Gamma  d x, \\
 d \tilde{p} = \Gamma  d p + \Xi  d x,
\end{array}
\end{equation}

\noindent that in component form are (with units where $c=1$)
\begin{equation}
\begin{array}{l}
 d {\tilde{x}}^{a} =\lambda _{b}^{a} d x^{b}, \\
 d {\tilde{p}}^{a} =\lambda _{b}^{a} d p^{b}+ \xi _{b}^{a} d x^{b}.
\end{array}
\end{equation}

\noindent Then, the proper time line element is invariant as required
by construction\ \ 
\begin{equation}
\begin{array}{rl}
 d \tau ^{2} & =\eta _{a,b}d {\tilde{x}}^{a}d {\tilde{x}}^{b}=\eta
_{a,b}\lambda _{c}^{a} d x^{c} \lambda _{d}^{b} d x^{d} \\
  & =\eta _{a,b}d x^{a}d x^{b}.
\end{array}
\end{equation}

The $\lambda _{c}^{a} $ are the components of the Lorentz transformation
that as usual depend on the relative rotation angle and hyperbolic
boost\ \ angle.\ \ The mass $\mu $ satisfies 
\begin{equation}
\begin{array}{rl}
 c^{2 }d {\tilde{\mu }}^{2} & =\eta _{a,b}d {\tilde{p}}^{a}d {\tilde{p}}^{b}
\\
  & =\eta _{a,b}( \lambda _{c}^{a} d p^{c}+\xi _{c}^{a} d x^{c})
\left(  \lambda _{d}^{c} d p^{d} + \xi _{d}^{b} d x^{d}\right) 
\\
  & =c^{2 }d \mu ^{2}+\eta _{a,b}\xi _{c}^{a}\xi _{d}^{b} d x^{c}d
x^{d}+2 \eta _{a,b}\xi _{c}^{a} \lambda _{d}^{b} d x^{c}d p^{d}.
\end{array}
\end{equation}

\noindent From basic dimensional analysis, the $\xi _{c}^{a}$ have
the dimensions of force or power (in units with $c=1$ these are
the same).\ \ It is a symmetric tensor satisfying $\xi _{b}^{a}=\eta
^{a,c}\eta _{b,d}\xi _{c}^{d}$ that transforms as an (1,1) tensor
under the Lorentz transformation 
\begin{equation}
{\tilde{\xi }}_{b}^{a}=\lambda _{c}^{a}\lambda _{b}^{d}\xi _{d}^{c}.
\end{equation}

These are the properties of a power-force stress tensor that is
the proper time derivative of the energy-momentum stress tensor.

\noindent The rate of change of the mass squared with respect to
the proper time is given by 
\begin{equation}
\frac{d {\tilde{\mu }}^{2}}{d \tau ^{2}}=\frac{d \mu ^{2}}{d \tau
^{2}}+\frac{1}{c^{2}}\eta _{a,b}\xi _{c}^{a}V^{c}( \xi _{d}^{b}
V^{d}+2\ \ \lambda _{d}^{b} F^{d}) 
\end{equation}

\noindent where $V^{a}=\frac{d x^{a}}{d \tau }$ is the {\itshape
four} velocity and $F^{a}= \frac{d p^{a}}{d \tau }$ is the {\itshape
four} force for the case $n=3$.\ \ \ 

\subsection{Three notation}

Further insight into the physical meaning of the group may be obtained
by converting to $n+1$ notation that for $n=3$ is the familiar {\itshape
three} notation $\{x^{a}\}=\{t,\frac{1}{c}q^{i}\}$, $\{p^{a}\}=\{\frac{1}{c}e,p^{i}\}$,
$i,j=1,..n$.\ \ The Lorentz matrix $\Lambda ( \alpha ,\beta ) $
parameterized by rotation angles $\alpha ^{i,j}=-\alpha ^{j,i}$
and hyperbolic boost rotations $\beta ^{i}$ that have the usual
form.\ \ For simplicity, we give here only the case $\alpha ^{i,j}=0$,
\begin{equation}
\Lambda ( 0,\beta ) = \left( \begin{array}{ll}
 \lambda _{0}^{0} & \lambda _{i}^{0} \\
 \lambda _{0}^{j} & \lambda _{i}^{j}
\end{array}\right) =\left( \begin{array}{ll}
 \cosh  \left( \beta \right)  & \sinh  \left( \beta \right)  \frac{
\beta _{i}}{c \beta }  \\
 c \sinh  \left( \beta  \right) \frac{\beta ^{j} }{\beta }\ \ \ 
& {\delta  }_{i}^{j} +\left(  \cosh ( \beta ) -1\right)  \frac{
\beta ^{j}\beta _{i}}{\beta ^{2}}
\end{array}\right) ,
\end{equation}

\noindent where $\beta ^{2}=\beta _{i}\beta ^{i}$. Indices are raised
and lowered with the kronecker delta $\delta _{i,j}$. As usual,
we identify velocity as $v^{i} =c \frac{\beta ^{i} }{\beta } \tanh
(\beta  ) $ and define\ \ $\gamma ( \beta ) = \cosh ( \beta ) =\lambda
_{0}^{0}$ or equivalently $\gamma ( v) ={(1-{(\frac{v}{c})}^{2})}^{-\frac{1}{2}}$.\ \ 

The velocity {\itshape four} vectors are given as usual by $\{V^{0},V^{i}\}=\{\gamma
,\gamma  v^{i}\}=\gamma \{1,\frac{d x^{i}}{d t}\}$ where\ \ $\gamma
=\frac{d t}{d \tau }$.\ \ The {\itshape four} force likewise is
$\{F^{0},F^{i}\}=\{\gamma  r,\gamma  f^{i}\}$ where $f^{i}=\frac{d
p^{i}}{d t}$ and $r=\frac{d e}{d t}$ and\ \ $f^{i}$ has the dimensions
of force and $r$ has the dimensions of power.\ \ The power-force-stress
components are
\begin{equation}
\Xi = \left( \begin{array}{ll}
 \xi _{0}^{0} & \xi _{i}^{0} \\
 \xi _{0}^{j} & \xi _{i}^{j}
\end{array}\right) =\gamma \left( \begin{array}{ll}
 \frac{1}{ c}r & -f_{i} \\
 f^{j} & \frac{1}{ c}m^{j,i}
\end{array}\right) .
\end{equation}

Therefore, the transformation equations for the position, time,
momentum, energy basis is 
\begin{equation}
\begin{array}{l}
 d\tilde{t}= \gamma  d t+\frac{1}{c} \lambda _{i}^{0}\ \ d q^{i},
\\
 d{\tilde{q}}^{i}=\lambda _{j}^{i}d q^{j} +c \lambda _{0}^{i}\ \ \ d
t, \\
 d{\tilde{p}}^{i}=\lambda _{j}^{i}d p^{j} +\lambda _{i}^{0}\ \ \ d
t, \\
 d\tilde{e}= \gamma  d e+c \lambda _{i}^{0}\ \ d p^{i}-\gamma  f_{i}
d q^{i}+c \gamma  r d t\ \ .
\end{array}%
\label{nirg: genral tx equations}
\end{equation}

\noindent For $n=1$ these are simply
\begin{equation}
\begin{array}{l}
 d\tilde{t}= \gamma ( v)  \left( \ \ d t+\frac{1}{c^{2}} v d q \right)
, \\
 d\tilde{q}=\gamma ( v) \left( d q +v d t \right) , \\
 d\tilde{p}=\gamma ( v) \left( d p^{j} +\frac{1}{c^{2}}v d e+f d
t+\frac{1}{c^{2}}\ \ m\ \ d q\right) , \\
 d\tilde{e}= \gamma ( v)  \left( d e+v d p-f d q+r d t \right) 
. 
\end{array}
\end{equation}

\noindent and the corresponding group parameter transformations
using (31) are 
\begin{gather}
\gamma ( v) \left( \begin{array}{ll}
 1 & v \\
 v & 1
\end{array}\right) =\gamma ( v^{{\prime\prime}}) \gamma ( v^{\prime
}) \left( \begin{array}{ll}
 1 & v^{\prime } \\
 v^{\prime } & 1
\end{array}\right) \left( \begin{array}{ll}
 1 & v^{{\prime\prime}} \\
 v^{{\prime\prime}} & 1
\end{array}\right) 
\end{gather}
\begin{equation}
\gamma ( v) \left( \begin{array}{ll}
 r & f \\
 f & m
\end{array}\right) =\gamma ( v^{{\prime\prime}}) \gamma ( v^{\prime
}) \left( \left( \begin{array}{ll}
 r^{\prime } & f^{\prime } \\
 f^{\prime } & m^{\prime }
\end{array}\right) \left( \begin{array}{ll}
 1 & v^{{\prime\prime}} \\
 v^{{\prime\prime}} & 1
\end{array}\right) +\left( \begin{array}{ll}
 1 & v^{\prime } \\
 v^{\prime } & 1
\end{array}\right) \left( \begin{array}{ll}
 r^{{\prime\prime}} & f^{{\prime\prime}} \\
 f^{{\prime\prime}} & m^{{\prime\prime}}
\end{array}\right) \right) 
\end{equation}

\noindent and so
\begin{equation}
\begin{array}{l}
 v=\left( v^{{\prime\prime}}+ v^{\prime } \right) /\left( 1+ \frac{v^{\prime
}v^{{\prime\prime}}}{c^{2}}\right) , \\
 f=\left( f^{{\prime\prime}}+f^{\prime } \mathrm{+}\frac{1 }{c^{2}}\left(
r^{\prime } v^{{\prime\prime}}-v^{\prime } r^{{\prime\prime}}\right)
\right) /\left( 1+ \frac{v^{\prime }v^{{\prime\prime}}}{c^{2}}\right)
, \\
 r=\left( r^{{\prime\prime}}\mathrm{+}r^{\prime }-f^{\prime } v^{{\prime\prime}}+v^{\prime
}f^{{\prime\prime}} \right) /\left( 1+ \frac{v^{\prime }v^{{\prime\prime}}}{c^{2}}\right)
. \\
 m=\left( m^{{\prime\prime}}\mathrm{+}m^{\prime }+f^{\prime } v^{{\prime\prime}}-v^{\prime
}f^{{\prime\prime}} \right) /\left( 1+ \frac{v^{\prime }v^{{\prime\prime}}}{c^{2}}\right)
.
\end{array}
\end{equation}

Consider next the algebra,\ \ we have the infinitesimal parameter
correspondence
\begin{equation}
\lambda ^{0,i}=\frac{1}{c}\beta ^{i}, \lambda ^{j,i}=\alpha ^{i,j},
\xi ^{0,0}=\frac{1}{ c}r,\xi ^{j,0}=f^{j}, \xi ^{j,i}=\frac{1}{
c}m^{j,i}
\end{equation}

\noindent where $\alpha ^{i,j}=-\alpha ^{i,j}$ and $m^{i,j}=m^{j,i}$
with the corresponding generators 
\begin{equation}
L_{0,j}=c K_{j}, L_{i,j}=J_{i,j}, M_{0,0}=c R, M_{i,0}=N_{i}, M_{i,j}=c
M \mbox{}^{\circ}_{i,j}.
\end{equation}

A general element of the algebra is\ \ 
\begin{equation}
Z= \alpha ^{i,j}J_{i,j}+\beta ^{i}K_{i}+f^{i}N_{i} +r R+ m^{i,j}M
\mbox{}^{\circ}_{i,j}.
\end{equation}

 The nonzero commutators of the Lie algebra (37) written in terms
of these generators is 
\begin{equation}
\begin{array}{l}
 \left[ J_{i,j},J_{k,l}\right] =-J_{j,l} \delta _{i,k}+J_{j,k} \delta
_{i,l}+J_{i,l} \delta _{j,k}-J_{i,k} \delta _{j,l}, \\
 \left[ J_{i,j},J_{k,l}\right] =-J_{j,l} \delta _{i,k}+J_{j,k} \delta
_{i,l}+J_{i,l} \delta _{j,k}-J_{i,k} \delta _{j,l}, \\
 \left[ J_{i,j},K_{k}\right] =-K_{j} \delta _{i,k}+K_{i} \delta
_{j,k},\ \ \ \left[ K_{i},K_{k}\right] =\frac{1}{c^{2}}J_{i,k},
\\
 \left[ J_{i,j},N_{k}\right] =-N_{j} \delta _{i,k}+N_{i} \delta
_{j,k},\ \ \ \left[ K_{i},N_{k}\right] =-M_{i,k}-R \delta _{i,k},
\\
 \left[ K_{i},R\right] =-\frac{2 }{c^{2}}N_{i}, \\
 \left[ J_{i,j},M \mbox{}^{\circ}_{k,l}\right] =M \mbox{}^{\circ}_{j,l}
\delta _{i,k}-M \mbox{}^{\circ}_{j,k} \delta _{i,l}+M \mbox{}^{\circ}_{i,l}
\delta _{j,k}+M \mbox{}^{\circ}_{i,k} \delta _{j,l}, \\
 \left[ K_{i},M \mbox{}^{\circ}_{k,l}\right] =\frac{-1}{c^{2}}\left(
N_{l} \delta _{i,k}+N_{k} \delta _{i,l}\right) .
\end{array}%
\label{nirg: Ub Algebra in three form}
\end{equation}

\subsection{Contraction in the limit $c\rightarrow \infty $}

 The scaling with $c$ given in (54) satisfies the conditions for
an In\"on\"u-Wigner contraction \cite{Glimore2} $c\rightarrow \infty
$ for which the nonzero contracted commutators are
\begin{equation}
\begin{array}{l}
 \left[ J_{i,j},J_{k,l}\right] =-J_{j,l} \delta _{i,k}+J_{j,k} \delta
_{i,l}+J_{i,l} \delta _{j,k}-J_{i,k} \delta _{j,l}, \\
 \left[ J_{i,j},K_{k}\right] =-K_{j} \delta _{i,k}+K_{i} \delta
_{j,k}, \\
 \left[ J_{i,j},N_{k}\right] =-N_{j} \delta _{i,k}+N_{i} \delta
_{j,k}, \\
 \left[ K_{i},N_{k}\right] =-M \mbox{}^{\circ}_{i,k}-R \delta _{i,k},
\\
 \left[ J_{i,j},M \mbox{}^{\circ}_{k,l}\right] =-M \mbox{}^{\circ}_{j,l}
\delta _{i,k}-M \mbox{}^{\circ}_{j,k} \delta _{i,l}+M \mbox{}^{\circ}_{i,l}
\delta _{j,k}+M \mbox{}^{\circ}_{i,k} \delta _{j,l}.
\end{array}
\end{equation}

 The subgroup spanned by $\{J_{i,j}, K_{i}, N_{i}, R\}$ is the algebra
of the Hamilton group $\mathcal{H}a( n) $.\ \ The full\ \ algebra
of the group that we call $\mathcal{U}bc( n) $ that is defined by
\begin{equation}
\mathcal{U}bc( n) = \mathcal{O}( n) \otimes _{s}\mathcal{A}( n(
n+1) /2) \otimes _{s}\mathcal{H}( n) 
\end{equation}

\noindent where $\{J_{i,j}\}$ are the generators of $\mathcal{O}(
n) $, $\{M_{i,j}\}$ are the generators of $\mathcal{A}( n( n+1)
/2) $ and $\{K_{i},N_{i},R\}$ are the generators of the Weyl-Heisenberg
group $\mathcal{H}( n) $.\ \ 

Furthermore, we take the limit\ \ $c\rightarrow \infty $ so that
$\beta \rightarrow 0$ in such a way that $c \beta =\tilde{\beta
}$ is finite, 
\begin{equation}
v=\operatorname*{\lim }\limits_{c\,\rightarrow \:\infty }c \tanh
\frac{\tilde{\beta }}{c}\ \ =\tilde{\beta },\ \ \ \ \ \ \operatorname*{\lim
}\limits_{c\,\rightarrow \:\infty } \gamma ( \frac{\tilde{\beta
}}{c}) =1
\end{equation}

\noindent The basis transformation equations (46) contract to\ \ the
expected transformation equations in the limit \cite{Low7}
\begin{equation}
\begin{array}{l}
 d\tilde{t}=\ \ d t, \\
 d{\tilde{q}}^{i}={\lambda ( \alpha ) }_{j}^{i}d q^{j} +v^{i}\ \ d
t, \\
 d{\tilde{p}}^{i}={\lambda ( \alpha ) }_{j}^{i}d p^{j} +f^{i}d t,
\\
 d\tilde{e}= d e+v_{i}d p^{i}-f_{i} d q^{i}+r d t. 
\end{array}
\end{equation}

\noindent where the ${\lambda ( \alpha ) }_{j}^{i}$ are now the
components of a rotation matrix, $\mathcal{O}( n) $. 

\subsection{Contraction from $\mathcal{U}( 1,n) $ of reciprocal
relativity in limit $b\rightarrow \infty $}

Similar considerations using the Born-Green metric 
\begin{equation}
d s^{2}= d x^{a}d x^{b }+\frac{1}{b^{2}}d p^{a}d p^{b}
\end{equation}

\noindent instead of the degenerate Minkowski line element results
in a reciprocal relativity theory where the homogeneous group is
$\mathcal{U}( 1,n) $ \cite{Low5},\cite{Low6}.\ \ This theory requires
the introduction of a constant $b$ with the dimensions of force
that may be taken to be on of the three universal constants that
form the natural dimensional basis $\{c,b,\hbar \}$.\ \ This is
instead of the usual choice $\{c,G,\hbar \}$ where $G$ is the gravitational
coupling constant. $G$ may be defined in terms of $b$ (or vice versa)
as $G=\alpha _{b}\frac{c^{4}}{b}$ where $\alpha _{b}$ is the dimensionless
gravitational coupling constant. 

The algebra for the $\mathcal{U}( 1,n) $ group is 
\begin{equation}
\begin{array}{l}
 \left[ L_{a,b},L_{c,d}\right] =-L_{b,d} \eta _{a,c}+L_{b,c} \eta
_{a,d}+L_{a,d} \eta _{b,c}-L_{a,c} \eta _{b,d}, \\
 \left[ L_{a,b},M_{c,d}\right] =-M_{b,d} \eta _{a,c}-M_{b,c} \eta
_{a,d}+M_{a,d} \eta _{b,c}+M_{a,c} \eta _{b,d}, \\
 \left[ M_{a,b},M_{c,d}\right] =-\frac{1}{b^{2}}\left( L_{b,d} \eta
_{a,c}+L_{b,c} \eta _{a,d}+L_{a,d} \eta _{b,c}+L_{a,c} \eta _{b,d}\right)
.
\end{array}%
\label{nirg: Ub algebra}
\end{equation}

This satisfies the condition for an In\"on\"u-Wigner contraction
to the algebra for the $\mathcal{U}b( 1,n) $ group given in (37)
in the limit $b\rightarrow \infty $. (This is where the notation
'$\mathcal{U}b$' originates.)

The $\mathcal{U}b( 1,n) $ group defines the limiting behavior of
reciprocal relativity in the limit of small interactions where the
systems are {\itshape almost inertial.\ \ }These are the {\itshape
expected} transformations from an analysis of noninertial frames
in a special relativistic context. 

\section{ Summary}

We started by noting that neither special relativity nor general
relativity address the problem of how clocks of noninertial states
due to forces other than gravity where gravity is negligible and
therefore the manifold is flat. 

The hypothesis that the Minkowski proper time line element is invariant
in these noninertial states (that includes the inertial states of
special relativity) and also requiring that the Heisenberg commutation
relations hold in all noninertial states results in the noninertial
relativity group $\mathcal{U}b( 1,n) $.\ \ This group give expected
transformations to noninertial states in terms of a power-force
stress tensor that is the proper time derivative of the energy-momentum
stress tensor.\ \ A general formula for the {\itshape non-quantum
classical} decay rate of mass for noninertial frames is derived.

The $\mathcal{U}b( 1,n) $ group is also the $b\rightarrow \infty
$ of the $\mathcal{U}( 1,n) $ group of reciprocal relativity described
in \cite{Low5},\cite{Low6}.\ \ This gives an understanding of the
behavior of reciprocal relativity in the small interaction limit
(that is, small forces relative to $b$)\ \ that is analogous to
the manner in which the Euclidean group that is the homogeneous
group of the Galilei group gives the small velocity limit, relative
to $c$, of the Lorentz group. 

Spacetime is an invariant subspace under the actions of the $\mathcal{U}b(
1,n) $ group and therefore is observer independent or absolute.
In this limit,\ \ there is an apparent global inertial frame that
all observers agree on.\ \ Forces appear to be relative to this
frame rather than being strictly relative to particle states. Forces
and the power-force-stress energy tensor are simply additive and
unbounded. Velocities are bounded by $c$ and strictly relative to
particle states. 

In the $c\rightarrow \infty $ limit yields the classical {\itshape
nonrelativistic} Hamilton theory that describes particles undergoing
general noninertial motion.\ \ \ In this case, there is an apparent
global inertial rest frame that all observers agree on.\ \ Forces
and velocities appear to be relative to this frame rather than being
strictly relative to particle states.\ \ Forces and velocities are
simply additive and unbounded. 

In a companion paper, the quantum mechanics that results from the
projective representations of the inhomogeneous $\mathcal{U}b( 1,n)
$ group for these noninertial states is studied using the same method
of projective representations of the inhomogeneous Lorentz group
for the inertial states of special relativistic quantum mechanics
\cite{wigner},\cite{Weinberg1}.

\section{Appendix A: Automorphisms of the Weyl-Heisenberg group
$\mathcal{H}( m) $}\label{nirg: Appendix A automorphisms of Heisenberg
group}

The Weyl-Heisenberg group $\mathcal{H}( m) $ has the matrix realization
as a subgroup of\ \ $\mathcal{G}\mathcal{L}( 2m+2) $ given by
\[
 \Upsilon ( z,\iota ) =\left( \begin{array}{lll}
 I_{2m} & 0 & z \\
  ^{t}\left( \zeta  z\right)   & 1 & \iota  \\
 0 & 0 & 1
\end{array}\right)  
\]

\noindent where $z\in \mathbb{R}^{2m}$, $\iota \in \mathbb{R}$.\ \ The
group multiplication and inverse are
\begin{equation}
\Upsilon ( z^{\prime },\iota ^{\prime }) \cdot \Upsilon ( z,\iota
) =\Upsilon ( z+z^{\prime },\iota +\iota ^{\prime }+z^{\prime }\cdot
\zeta  \cdot z) ,\ \ {\Upsilon ( z,\iota ) }^{-1}=\Upsilon ( -z,-\iota
) .
\end{equation}

$\mathcal{H}( m) $ has a group manifold diffeomorphic to $\mathbb{R}^{2m+1}$
and is therefore simply connected and is its own cover, $\overline{\mathcal{H}}(
m) \simeq \mathcal{H}( m) $, 

Elements $\Omega $ of the linear automorphism group ${\mathrm{aut}}_{\mathcal{H}(
m) }\subset \mathcal{G}\mathcal{L}( 2m+2) $\ \ that is a matrix
group that may be represented by $(2m+2)\times (2m+2)$ nonsingular
matrices $\Omega $ that satisfy 
\begin{equation}
\Omega  \Upsilon ( w^{\prime },\iota ^{\prime })  {\Omega }^{-1}=\Upsilon
( w^{{\prime\prime}},\iota ^{{\prime\prime}}) .%
\label{nirg: automorphism condition}
\end{equation}

The proof given by Folland \cite{folland} shows that the most general
matrix group with this property is $\Omega \in \mathcal{A}_{\mathcal{H}}$
where
\begin{equation}
\mathcal{A}_{\mathcal{H}}\simeq \mathbb{Z}_{2}\otimes _{s}\mathcal{D}\otimes
_{s}\mathcal{H}\mathcal{S}p( 2n+2) .%
\label{nirg: autH definition}
\end{equation}

 This can be shown by direct matrix computation that the most general
elements of\ \ $\mathcal{G}\mathcal{L}( 2m+2) $ satisfying (62)
are 
\begin{equation}
\Omega ( \epsilon ,\delta ,\Sigma ,z,\iota ) =\left( \begin{array}{lll}
 \delta  \Sigma  & 0 & z \\
 -^{t}z \zeta  \Sigma  & \epsilon  \delta ^{2} & \iota  \\
 0 & 0 & \epsilon 
\end{array}\right)  ,%
\label{matrix reps of AutH}
\end{equation}

\noindent where $A\in \mathcal{S}p( 2m) $, $z\in \mathbb{R}^{2m}$,
$\delta ,\iota  \in \mathbb{R}$, $\epsilon =\pm 1$\ \ and $\zeta
$ is the $2m\times 2m$ symplectic matrix. The group multiplication
and inverse are
\begin{equation}
\begin{array}{l}
 \begin{array}{rl}
 \Omega ( \epsilon ^{{\prime\prime}},\delta ^{{\prime\prime}},\Sigma
^{{\prime\prime}},z^{{\prime\prime}},\iota ^{{\prime\prime}})  &
=\Omega ( \epsilon ,\delta ,\Sigma ,z,\iota ) \Omega ( \epsilon
^{\prime },\delta ^{\prime },\Sigma ^{\prime },z^{\prime },\iota
^{\prime }) 
\end{array} \\
 =\Omega ( \epsilon  \epsilon ^{\prime },\delta  \delta ^{\prime
},\Sigma  \Sigma ^{\prime },\epsilon ^{\prime }z+\delta  \Sigma
z^{\prime },\epsilon ^{\prime }r+\epsilon  \delta ^{2}r^{\prime
}{- }^{t}z \zeta  \Sigma  z^{\prime } )  \\
 {\Omega ( \epsilon ,\delta ,\Sigma ,z,r) }^{-1}=\Omega ( \epsilon
,\delta ^{-1},\delta ^{-1}\Sigma ^{-1},-\epsilon  \delta ^{-1}\Sigma
^{-1}z,-\delta ^{-2}r) 
\end{array}
\end{equation}

\noindent Note that
\begin{equation}
\begin{array}{l}
 \Omega ( 1,1,I_{2n},z,\iota ) \simeq \Upsilon ( z,\iota ) \in \mathcal{H}(
m)  \\
 \Omega ( 1,1,\Sigma ,0,0) \simeq \Sigma \in \mathcal{S}p( 2n) 
\\
 \Omega ( \epsilon ,\delta ,1,0,0) \simeq \Delta ( \epsilon ,\delta
) \in \mathcal{D}\otimes \mathbb{Z}_{2}
\end{array}
\end{equation}

\noindent with $\mathrm{Det} \Omega =\delta ^{2m+2}$ where 
\begin{equation}
\Delta ( \epsilon ,\delta ) =\left( \begin{array}{lll}
 \delta   & 0 & 0 \\
 0 & \epsilon  \delta ^{2} & 0 \\
 0 & 0 & \epsilon 
\end{array}\right)  ,\ \ \Sigma \simeq \left( \begin{array}{lll}
 \Sigma   & 0 & 0 \\
 0 & 1 & 0 \\
 0 & 0 & 1
\end{array}\right)  .%
\label{matrix reps of AutH}
\end{equation}

The automorphism group may be written as
\begin{equation}
\Omega ( \epsilon ,\delta ,\Sigma ,\tilde{z},\tilde{\iota }) =\Delta
( \epsilon ,\delta ) \Sigma  \Upsilon ( z,\iota ) 
\end{equation}

\noindent where 
\begin{equation}
\tilde{z}= \delta  \Sigma  z,\ \ \tilde{\iota }= \epsilon  \delta
^{2}\iota .
\end{equation}

The above discussion gives the automorphism group $\mathcal{A}_{\mathcal{H}}$
that is a matrix group through direct computation of (62) with the
matrix group $\mathcal{H}( m) $.\ \ The central extension of this
group also defines automorphisms of $\mathcal{H}( m) $. This is
true because these elements are in the center of the group and as
they commute with all elements, they always satisfy (8).\ \ The
central extension of a matrix group is not necessarily a matrix
group \cite{Hall}.\ \ This is true in particular for $\mathcal{A}_{\mathcal{H}}$
which is why the matrix calculation does not give the central elements.
Therefore, to obtain the full automorphism group, we must calculate
the central extension\ \ ${\mathcal{A}ut}_{\mathcal{H}}={\check{\mathcal{A}}}_{\mathcal{H}}$.

The method of determining the central extension is given in \cite{Azcarraga},\cite{Weinberg1},\cite{Low8}.
It first requires the determination of the algebraic central extension
of the Lie algebra. Using the methods given in these references,
it may be shown that the algebra of ${\mathrm{aut}}_{\mathcal{H}}$
does not have a central extension. 

\noindent Therefore, the central extension is simply the universal
cover of the group. $\mathcal{D}$ has a group manifold diffeomorphic
to $\mathbb{R}$ and\ \ $\mathcal{H}( m) $ has a group manifold diffeomorphic
to $\mathbb{R}^{2m+1}$.\ \ The fundamental homotopy group for the
symplectic group is the integers under addition and so $\mathcal{S}p(
m) \simeq \overline{\mathcal{S}p}( m) /\mathbb{Z}$.\ \ \ Therefore,
\begin{equation}
\begin{array}{rl}
 {\mathcal{A}ut}_{\mathcal{H}( m) } & \simeq {\check{\mathcal{A}}}_{\mathcal{H}(
m) }\simeq {\overline{\mathcal{A}}}_{\mathcal{H}( m) } \\
  & =\left( \mathbb{Z}_{2}\otimes \mathcal{D}\otimes \overline{\mathcal{S}p}(
2m) \right) \otimes _{s}\mathcal{H}( m) 
\end{array}
\end{equation}

\noindent This issue of the central extensions is important for
the quantum mechanical treatment where the projective representations
are required.

\end{document}